\documentclass[final,5p,times,twocolumn,letterpaper]{elsarticle}

\usepackage{natbib}
\usepackage{cancel}
\usepackage{pifont} 
\usepackage{amsthm}         
\usepackage{amsmath} 	   
\usepackage{amssymb}       
\usepackage{amsfonts}
\usepackage{graphicx} 	    
\usepackage{verbatim}
\usepackage{epsfig,bbm}
\usepackage{color}
\usepackage{colortbl}
\usepackage{float}
\usepackage{multirow}
\usepackage{tensor}
\usepackage{amsmath}
\newenvironment{psmallmatrix}
  {\left(\begin{smallmatrix}}
  {\end{smallmatrix}\right)}

\usepackage{subcaption}

\definecolor{babyblue}{rgb}{0.54, 0.81, 0.94}
\definecolor{corn}{rgb}{0.98, 0.93, 0.36}

\begin{document}

\begin{frontmatter}

\title{The Effects of Multiple Modes and Reduced Symmetry on the Rapidity and Robustness of Slow Contraction}

\author[add1,add3]{Anna Ijjas}
\author[add0]{Frans Pretorius}
\author[add0]{Paul J. Steinhardt}
\author[add0]{Andrew P. Sullivan}

\address[add1]{Max Planck Institute for Gravitational Physics (Albert Einstein Institute), 30167 Hannover, Germany}
\address[add3]{Institute for Gravitational Physics, Leibniz University, 30167 Hannover, Germany}
\address[add0]{Department of Physics, Princeton University, Princeton, NJ 08544, USA}

\date{\today}

\begin{abstract}
We demonstrate that the rapidity and robustness of slow contraction in homogenizing and flattening the universe found in simulations in which the initial conditions were restricted to non-perturbative variations described by a single fourier mode along only a single spatial direction are in general {\it enhanced} if the initial variations are along two spatial directions, include multiple modes, and thereby have reduced symmetry.  Particularly significant are shear effects that only become possible when variations are allowed along two or more spatial dimensions.   Based on the numerical results, we conjecture that the counterintuitive enhancement occurs because
 more degrees of freedom are activated which drive spacetime away from an unstable Kasner fixed point and towards the stable Friedmann-Robertson-Walker fixed point. 
 \end{abstract}

\begin{keyword}
slow contraction, homogeneity, isotropy, mode coupling, numerical relativity, bouncing/cyclic cosmology 
\end{keyword}

\end{frontmatter}

{\it Introduction.} Slow contraction has been proposed as a mechanism for explaining the homogeneity, isotropy and spatial flatness of the universe \cite{Khoury:2001wf,Erickson:2006wc}.  During this phase, the contraction is `slow' because the Friedmann-Robertson-Walker (FRW) scale factor $a(\tau)$ decreases as a small power of FRW time $\tau$,
\begin{equation}
 a(\tau) \propto (-\tau)^{1/\varepsilon} \quad {\rm as} \quad \tau \rightarrow 0^{-},
 \end{equation}
where 
\begin{equation}
\varepsilon \equiv {\textstyle {\frac32}} \left(1 + {\textstyle {\frac{p}{\varrho}}} \right)  \gg 3 
\end{equation}
is the equation of state, $p$ is the pressure and $\varrho$ is the energy density of the dominant stress-energy component.   In a bouncing or cyclic cosmology based on slow contraction, the evolution connects to the hot expanding universe through a classically-described smooth transition (the `bounce')  and the conversion of energy driving slow contraction into hot matter and radiation.   

In order for a mechanism to explain why the universe is homogeneous, isotropic and spatially flat and not otherwise, it must pass several tests (see, {\it e.g.}, Ref.~\cite{Cook:2020oaj}). First, the mechanism must have the property that the relative contribution of small initial inhomogeneities and anisotropies to the total energy density shrink with time according to the classical cosmological evolution equations.  Second, the smoothing must occur when all quantum fluctuations are included.   For example, there should not be a quantum runaway that can transform spacetime, even an initially smooth spacetime, into a multiverse in which the fraction of spacetime that is smoothed is either exponentially small or indeterminate.  It is straightforward to show with pencil and paper that slow contraction passes both of these tests.  

The more challenging tests are to determine if the smoothing mechanism is robust, meaning effective even when initial conditions correspond to large, non-perturbative deviations from homogeneity, isotropy and spatial flatness (for simplicity, henceforth denoted by `variations'); and to determine if the mechanism is  rapid enough for the smoothing phase to endure long after a flat FRW spacetime is reached, sufficient for quantum fluctuations on the now-smoothed background to evolve into the density perturbations needed to account for the observed temperature variations in the cosmic microwave background and seed the formation of large-scale structure. Since the initial deviations from a flat FRW spacetime are necessarily non-perturbative, numerical relativity techniques adapted to cosmological backgrounds are required to perform these tests. 

Numerical relativity simulations demand significant computational resources. Hence, the first studies of robustness and rapidity only considered the simplest settings, {\it i.e.}, cases in which the initial variations of the energy density, curvature and shear were along a single spatial direction and described by a single fourier mode \cite{Garfinkle:2008ei,Ijjas:2020dws}.   The simulations showed that slow contraction exceeded the conditions for a rapid and robust smoother for $\varepsilon \gtrsim 13$.  These results are promising yet not sufficient.

In this paper, we complete the test using  simulations that allow variations along at least two spatial directions and the inclusion of two or more modes along each of those directions.  Given the inherently non-linear and non-perturbative nature of the evolution equations, this is essential to address the legitimate concern that there could be mode-coupling or multi-dimensional effects or  other aspects of initial conditions with reduced symmetry that inhibit smoothing and flattening which were missed in the earlier simulations.  As described below, what we find instead is the opposite.  The simulations indicate that, in general, the robustness and rapidity of slow contraction are enhanced by initial conditions with reduced symmetry.  We  conclude by commenting on why this might occur.

{\it Numerical Scheme and Initial Conditions.} 
Our simulations solve the full 3+1 dimensional coupled Einstein-scalar field equations beginning from large variations in spatial curvature, matter density and shear along two spatial directions and track their evolution for several hundreds of $e$-folds.     

 The scalar field is introduced as a microphysical source for achieving the desired macroscopic equation of state, $\varepsilon \gg 3$, characteristic of slow contraction.   The scalar field $\phi(\vec{x},t)$ can vary with space and time and evolves along  a negative exponential potential $V(\phi) = V_0 e^{ \phi/m}$, where $V_0 <0$. Throughout, we express microphysical quantities in reduced Planck units, $8 \pi G_N=1$ with $G_N$ being Newton's constant.  In the homogeneous limit, the  attractor equation of state $\varepsilon^{\ast}$ associated with the scalar field $\phi$ is given by  $m= 1/\sqrt{2 \varepsilon^{\ast}}$. Large values of  $\varepsilon^{\ast}$ can be reached with values of $m$ that are modestly smaller than one; for example, $m=0.1$ induces a slow contraction phase with $\varepsilon^{\ast}= 50$.  

Earlier studies \cite{Cook:2020oaj,Garfinkle:2008ei,Ijjas:2020dws} based on initial variations along a single spatial direction studied the phase space for smoothing for a wide range of $\varepsilon^{\ast}$ and initial states.  The results  showed that slow contraction exceeds the conditions for rapidity and robustness for $\varepsilon^{\ast} \gtrsim 13$.  In order to determine whether the inclusion of variations along two spatial dimensions and with multiple modes in each direction disfavors (or favors) smoothing, our examples use the minimal value $\varepsilon^{\ast}=13$ or $m=0.196$.

 The numerical scheme for evolving the corresponding system of coupled, non-linear partial differential equations is based on a Hubble-normalized, orthonormal tetrad formulation of the Einstein-scalar field equations which is fully detailed in Ref.~\cite{Ijjas:2021gkf}. 
 For example, we define the scalar field time derivative 
\begin{equation}
\label{wdef}
\bar{W} = {{\cal N}^{-1}} {\partial _t} \phi ,
\end{equation}
where ${\cal N} = N/\Theta$ is the scale-invariant generalization of the lapse $N$, and
the time coordinate $t$ is given through
\begin{equation}
e^t = {\textstyle \frac13} \Theta,
\label{timechoice}
\end{equation}
with $\Theta = | H^{-1} |$ being the inverse mean curvature of constant time hypersurfaces, such that surfaces of constant time are constant mean curvature hypersurfaces. 

With Hubble-normalization, the variables tracked in the simulation correspond in the homogeneous limit to the dimensionless Friedmann variables, $\Omega_i$,  the fractional contribution of component $i$ (matter density, curvature or anisotropy) to  $H^2$ in the Friedmann constraint, see {\it e.g.} Ref.~\cite{Ijjas:2020dws}.  Note also that the matter contribution ($\Omega_m$), which includes the sum of positive kinetic energy density and negative potential energy density, and the curvature contribution  ($\Omega_k$) can be positive or negative. In addition, employing Hubble-normalized variables enables running the simulation for many $e$-folds of contraction of the Hubble radius without reaching the putative singularity or encountering stiff behavior. 

To specify the spatial hypersurface with mean curvature $\Theta_0^{-1}$ at some initial time $t_0$, we adapt the York method \cite{York:1971hw} commonly used in numerical relativity computations:
We define the spatial metric of the $t_0$-hypersurface to be conformally-flat and freely choose the vacuum contribution  $\tensor*{Z}{_a_b^0}(\vec{x},t_0)$
of the conformally rescaled trace-free extrinsic curvature,
\begin{equation}
\tensor*{Z}{_a_b}(\vec{x},t_0) \equiv \psi ^{6}(\vec{x},t_0) \bar{\Sigma} _{ab}(\vec{x},t_0),
\end{equation}
which is related to  the Hubble-normalized initial shear $\bar{\Sigma}_{ab}(\vec{x},t_0)$ through a conformal factor 
$\psi(\vec{x},t_0)$ set by the Hamiltonian constraint.  The momentum constraint determines the 
remainder of $\tensor*{Z}{_a_b}(\vec{x},t_0)$. 
Our choice of initial geometry enables us to freely specify the initial field $\phi(\vec{x},t_0)$ and velocity distributions 
\begin{equation}
\bar{W}(\vec{x},t_0) \equiv \psi ^{-6}(\vec{x},t_0) Q(\vec{x},t_0).
\end{equation}
Note that our choice of initial conditions is only restricted by the constraint equations of general relativity and the particular boundary conditions used in our simulations.

An example of a divergence-free and trace-less, conformally rescaled shear tensor $\tensor*{Z}{_a_b^0}(\vec{x},t_0)$ is given by
\begin{equation}
\tensor*{Z}{_a_b^0}(\vec{x},t_0) = 
\begin{psmallmatrix}
b_2 +  c_2 \cos{y} & {\;} & \xi & {\;} &  \kappa_1+ c_1 \cos{y}  \\
&&&&\\
\xi & \; &  b_1+  a_1  \cos{x} & {\;} &  \kappa_2 +a_2 \cos{x}  \\
&&&&\\
\kappa_1+ c_1 \cos{y} 
&  {\;} &  \kappa_2+ a_2 \cos{x}  &  {\;} & -b_1 - b_2 - a_1 \cos{x} - c_2 \cos{y}
\end{psmallmatrix},
\label{ZZ}
\end{equation}
where $a_1, a_2, b_1, b_2, c_1, c_2, \kappa_1, \kappa_2$ are constants and $\vec{x}= (x, \, y)$; in considering cases with multimodes, the cosine terms in the expression for $\tensor*{Z}{_a_b^0}(\vec{x},t_0)$ can be replaced by a sum of different fourier modes with different amplitudes, wavenumbers and phases that preserve the divergence and trace. We fix the initial scalar field velocity 
to be
\begin{equation}
\label{QQ}
Q(\vec{x},t_0)  = \Theta_0  \Big(q_{x} \cos{(m_x x + d_x)} + q_{y} \cos{(m_y y + d_y)} + Q_0 \Big) ,
\end{equation}
where  $Q_0$ is the mean value of the initial field velocity and 
$ q_x, q_y, m_x, m_y,  d_x, d_y$ are constant and denote  the amplitude, the mode number and the phase of the initial  velocity.  Here we show two modes but they can be replaced by a sum of different modes when considering cases with multimodes.    For both choice $\tensor*{Z}{_a_b^0}$ and $Q$,  expressing the spatial variation in fourier modes reflects the fact that there are periodic boundary conditions $0\leq x,y\leq2\pi$ with $0$ and $2\pi$ identified.  

Note that generality is not lost by choosing a conformally flat metric or choosing the scalar field to be uniform,  $\phi_0=0$, say, on the initial spatial hypersurface.  These are simply devices  to ensure constraint satisfying initial conditions and to simplify the specification of the initial conditions.  Constraint satisfaction propagates forward in time, as required, but conformal flatness and uniform $\phi$ are strongly broken in just a few integration steps.  

Two key parameters for determining whether spacetime ultimately converges to the slow contraction, smooth attractor described by a flat FRW geometry is the equation of state of the attractor solution, $\varepsilon=\varepsilon^{\ast}$, and the initial mean field velocity, $Q_0$ where $Q_0>0$ corresponds to the mean initial velocity along the potential $V(\phi)$ being downhill.  (N.B. In cyclic cosmologies based on slow contraction, the field velocity at the beginning of the contraction phase is in general downhill and large, see, {\it e.g.}, Ref.~\cite{Ijjas:2019pyf}.  However, as exemplified below, we also consider cases with uphill initial conditions as a method for understanding how and the degree as to which  slow contraction smooths and flattens spacetime.)  

In the case of homogeneous initial conditions, $\varepsilon^{\ast}$ and $Q_0$ are the only relevant parameters. For a given $\varepsilon^{\ast}$
and uniform initial field velocity $Q(\vec{x}, t_0)= Q_0$ for all $\vec{x}$, there is a critical initial value  $Q_0^{\ast}$ such that the universe converges to the slow contraction attractor solution for $Q_0 > Q_0^{\ast}$ and converges to a Kasner-like anisotropic solution for $Q \le Q_0^{\ast}$.  In the first case, both the scalar field kinetic and potential energy density rapidly grow to dominate, driving the universe towards a spatially flat FRW spacetime with $Q(\vec{x}, t)\rightarrow Q_{\rm attr}(\varepsilon^{\ast}, t)$, the attractor solution for the field velocity. In the second case, the scalar field potential remains irrelevant and the universe approaches a homogeneous anisotropic (`Kasner-like') phase with a mix of uniform shear and uniform scalar field kinetic energy density.  

\begin{figure}[t!]
\begin{center}
\includegraphics[width=1.90in,angle=-0]{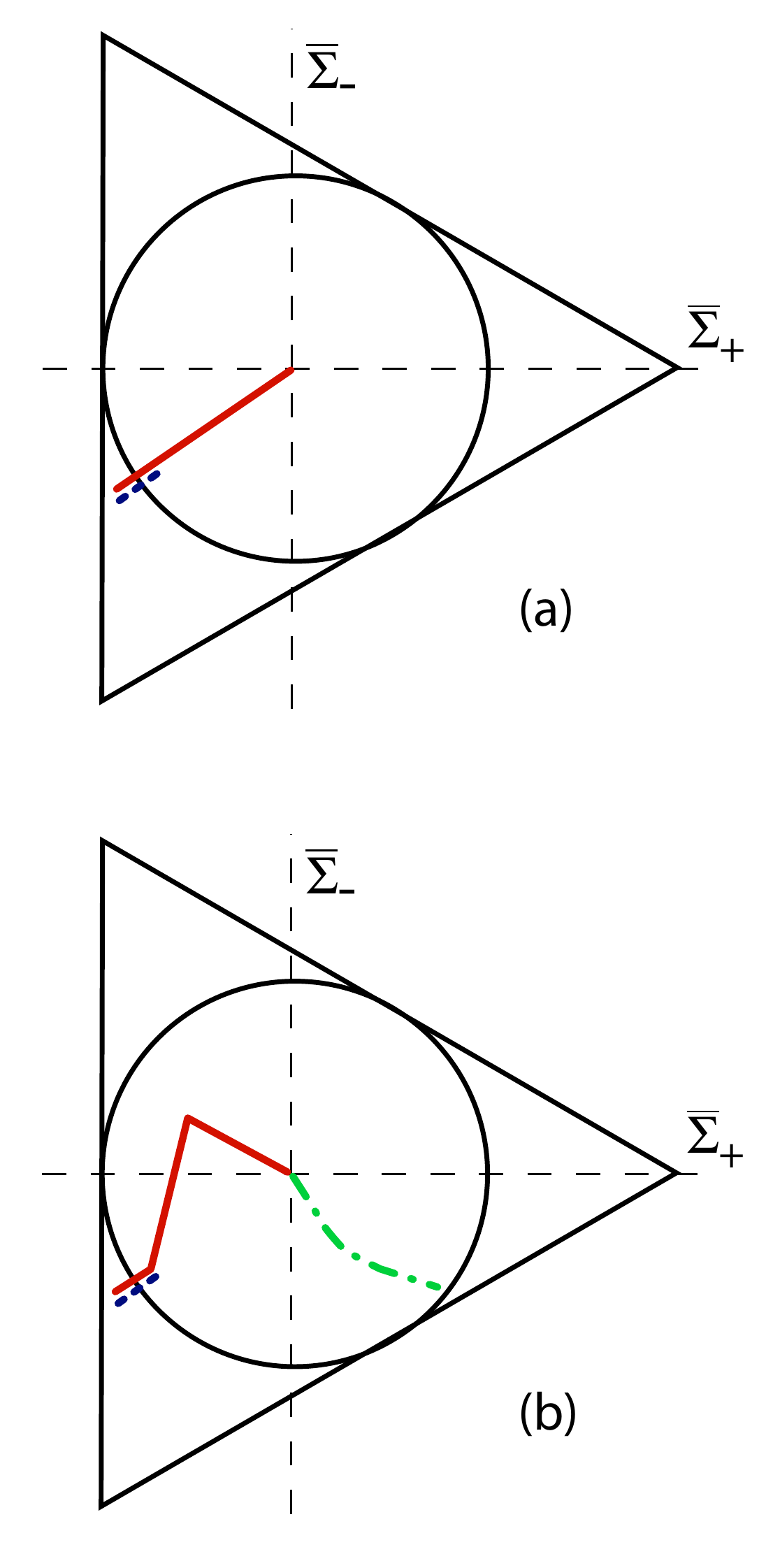}
\end{center}
\caption{The state space orbit for worldlines at
$(x=0.9 \pi,\, y=1.4 \pi)$ (though the final state does not depend on $(x,\,y)$ in these examples).  Fig. 1(a) shows two test cases with  homogeneous initial conditions for $\varepsilon^{\ast}=13$: $Q_0=Q_0*=0$ (short dashed blue), the critical value for converging to a Kasner-like anisotropic solution (near the Kasner circle); and $Q_0=0.001$ (solid red), a slightly downhill initial scalar field velocity, which instead evolves to the  flat FRW attractor solution (center of the circle).   Figure 1(b) shows three cases with non-uniform initial conditions that evolve to uniform spacetimes.  The short dashed (blue) trajectory  is the result of a simulation
in which the initial $x$-dependent non-uniformity is purely along  $Z_{23}$ ($a_2=0.01$). Even at $t=-100$, the trajectory remains 
close the Kasner circle.  The red solid trajectory is a 
simulation with the same initial conditions except for a small added $y$-dependent off-diagonal mode $Z_{13}$ with ($c_1=0.0001$); the example demonstrates that even a small shear in two dimensions is sufficient to kick the trajectory away from Kasner-like and drive it to the  flat FRW attractor solution.  The third case (dash-dotted, green) is a more general example with larger amplitude initial multi-mode spatial variations in the initial scalar field velocity and both diagonal and non-diagonal shear components.  The combination leads to a more complex trajectory that even more rapidly converges to the flat FRW attractor solution.
\label{fig:1}}
\end{figure}  

Figure~\ref{fig:1}a shows  state space orbit plots comparing the evolution beginning from homogeneous equations for 
$Q_0^{\ast}=0$ and a slightly positive value,  $Q_0=0.001$.  The orbit plots enable the visualization of the evolution of the shear  in the
$(\bar{\Sigma}_+ ,\bar{\Sigma}_-)$ plane for a chosen spatial point $x$, where
\begin{equation}\label{Sigmaplus}
\bar{\Sigma}_+ = \textstyle{\frac12}\Big(\bar{\Sigma}_{11} + \bar{\Sigma}_{22}\Big)
,\quad
\bar{\Sigma}_- = \textstyle{\frac{1}{2\sqrt{3}}}\Big(\bar{\Sigma}_{11} - \bar{\Sigma}_{22}\Big).
\end{equation}
The $\bar{\Sigma}_{\pm}$ are normalized so that the unit circle ($\bar{\Sigma}_+^2+\bar{\Sigma}_-^2 =1$) corresponds to the vacuum Kasner solution, as occurs for $Q_0=Q_0^{\ast}= 0$; trajectories that approach $\Omega_m=1$ (FRW)  converge to the center, as occurs for $Q_0=0.001$.  

When the special homogeneous initial conditions are replaced with non-perturbative, spatially varying initial conditions, the competition remains between the FRW and Kasner-like phases, but additional parameters  become relevant 
which lead to other outcomes.  Earlier studies based on numerical relativity codes that allow initial conditions with variations along one spatial direction only show that the outcome can be purely FRW, purely Kasner-like, or a mix with exponentially more proper volume in the FRW phase and the remainder in a Kasner-like phase that may be homogeneous or an unsmoothed phase that is both spatially and time varying with spiking \cite{Garfinkle:2008ei,Ijjas:2020dws}. However, for the overwhelming majority of the $\varepsilon-Q_0$ parameter space in which the average initial field velocity is downhill directed, the entire spacetime converges to the purely FRW slow contraction attractor phase, as desired in bouncing cosmology and cyclic models based on slow contraction \cite{Cook:2020oaj}.  

\begin{figure*}[t]
\begin{center}
\includegraphics[width=7.00in,angle=-0]{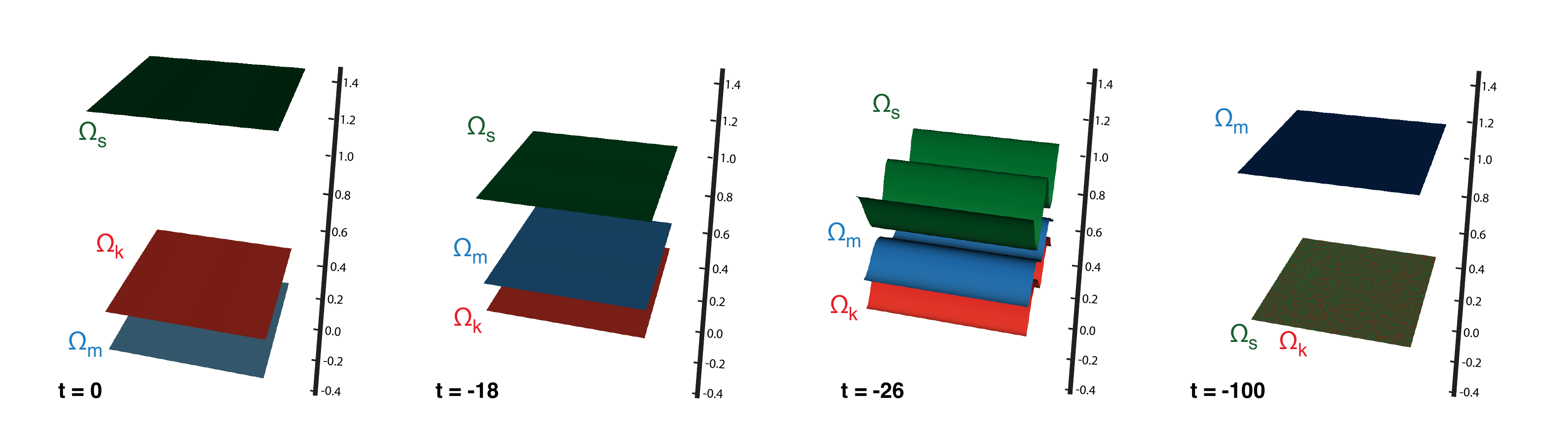}
\end{center}
\caption{The evolution of the relative contributions of the three energy density components, $\Omega_m=$~matter (blue), $\Omega_k=$~spatial curvature (red) and $\Omega_s=$~shear (green) for initial conditions in which the scalar field velocity and most components of  $\tensor*{Z}{_a_b^0} $ in Eq.~ (\ref{ZZ}) are spatially uniform except for an $x$-dependent fourier contribution to $\tensor*{Z}{_2_3^0} $ and $\tensor*{Z}{_3_2^0} $ with
  $a_2=0.01$ combined with a very small amplitude $y$-dependent contribution to $\tensor*{Z}{_1_3^0} $ and $\tensor*{Z}{_3_1^0} $ with  $c_1=0.0001$.
 Note that at $t=0$ the  shear  (green) dominates and matter and spatial curvature contributions are subdominant and that the universe converges to $\Omega_m=1$ flat FRW (and $\Omega_s=\Omega_k=0$) by $t=-100$. }    
\label{fig:2}
\end{figure*}

Here the goal is to use a more advanced numerical relativity code that allows initial conditions with variations along two independent spatial directions, reduced symmetry and  several variations along each spatial direction described by 
a sum of fourier modes to study whether rapidity and robustness are reduced or enhanced.  Our approach is to choose the smallest value of $\varepsilon^{\ast}$ that smoothed rapidly and robustly enough according to the earlier studies ($\varepsilon^{\ast}=13$); and then to choose
 the corresponding value of $Q_0=Q_0^{\ast}$ (which is $Q_0^{\ast}=0$ for this value of $\varepsilon^{\ast}$), so that the system is (slightly) biased towards not smoothing as far as these parameters are concerned.   
 That is, if the initial conditions were perfectly uniform, the spacetime would never evolve to the flat FRW state, as the short-dashed trajectory in 
 Fig.~\ref{fig:1}a shows.  Then we add to the initial conditions spatial variations that include   multiple modes and reduced symmetry to see if the spacetime is generally driven toward or away from the Kasner-like solution.  Note that $Q_0=Q_0^{\ast}$ and any variations we add to it are for {\it all} $\vec{x}$ very distant from the slow contraction attractor solution; {\it i.e.}, the initial conditions are in the highly non-perturbative regime far from a stable flat FRW phase.
 
If the spacetime is driven away from the Kasner-like solution and towards the flat FRW solution, we can conclude that multiple modes and/or reduced symmetry do no harm to or  even enhance the rapidity and robustness of smoothing through slow contraction.   
Similarly, we have also performed tests in which $Q_0$ is slightly biased towards the FRW attractor solution ($Q_0 > Q_0^{\ast}$).  If multiple modes and/or reduced symmetry drive the universe away from the flat FRW solution, we can conclude that they inhibit rapidity and robustness. 
In each case, we studied a wide range of non-perturbative spatially-varying initial conditions to determine what is the generic outcome.

\vspace{0.1in}
\noindent
{\it Results.}  We considered rapidity and robustness for three classes of initial conditions:

{\bf Multi-mode variations of the initial scalar field velocity $Q(\vec{x}, t_0)$ only.}
We started with initial conditions involving various combinations of fourier modes along  the $x$ and the $y$ directions for $Q(\vec{x}, t_0)$ but no spatial variations in the divergence-free part of the  shear $\tensor*{Z}{_a_b^0}(\vec{x}, t_0)$.  
Analogous earlier studies based on a single mode in $Q(\vec{x}, t_0)$ along one spatial direction concluded that the only relevant criterion deciding whether spacetime regions converge to the flat FRW slow contraction attractor solution is the degree to which the initial value of $Q(\vec{x}, t)$ is downhill in the vicinity of $\vec{x}$ or uphill; see Fig.~3 of Ref.~\cite{Ijjas:2020dws}.

In the new studies, we included two or more different cosine-like fluctuation modes about $Q_0^{\ast}$ that create peaks and valleys in  $Q(\vec{x}, t_0)$ across the two-dimensional $x-y$ plane.   The results are the straightforward generalization to two spatial dimensions of what was found for initial conditions that only varied along one spatial direction. Regions of the $x-y$ plane with $Q_0>Q_0^{\ast}$ generically evolve to the smooth, flat, FRW slow contraction attractor phase and regions with $Q_0<Q_0^{\ast}$ generically do not.  Furthermore, in the cases where the outcome is the smooth, flat FRW state, there was no detectable effect that the non-linear coupling of multiple modes and the reduced symmetry of the initial conditions had on the rapidity or robustness found in early studies with simpler initial conditions.  (In this and the studies described below, we verified numerical convergence using the protocols described in the appendix  of Ref.~\cite{Ijjas:2021gkf}.)

{\bf Initial shear variations only.}
More interesting results were found for initial conditions with spatially varying shear characterized by $\tensor*{Z}{_a_b^0}(\vec{x}, t_0)$ as given in Eq.~\eqref{ZZ} but with no perturbations of the initial scalar field velocity.   Early studies restricted to initial variations along a single spatial dimension activated a reduced set of shear components compared to what is possible in the new studies.

As an important instructive example, consider a case where the only spatially varying components of  the shear  $\tensor*{Z}{_a_b^0}(\vec{x}, t_0) $
are $\tensor*{Z}{_2_3^0}$ and $\tensor*{Z}{_3_2^0}$; they are purely $x$-dependent and have amplitude $a_2=0.01$.  
The remaining coefficients in Eq.~\eqref{ZZ}  are zero except for $b_2=1.8$, $b_1=-0.15$, and $b_3=-1.65$. 
In this case, the geometry approaches and remains close to a Kasner-like state even after  $t=-100$ (corresponding to  $e$-folds of contraction of the Hubble radius), as indicated by the fact that 
the corresponding state space orbit  illustrated in Fig.~\ref{fig:1} remains close to the Kasner circle.   Note that $Q_0=Q_0^{\ast}=0$ and $\varepsilon^{\ast}=13$, which slightly biases the outcome towards the Kasner-like state. The evolution reaches a uniform state with  $\Omega_s= 0.77$, $\Omega_m=0.23$ and $\Omega_k=0$ at $t=-100$.  The fact that the shear dominates and the spatial curvature is zero is significant because a burst of non-zero spatial curvature (a chaotic mixmaster type `bounce') is needed for the state to change to another Kasner-like state or the flat FRW slow contraction attractor state, and the spatial curvature appears to be inactive in this example.

Now, a single $y$-dependent, fourier mode with a very small amplitude ($c_1=0.0001$) is added to the off-diagonal components $\tensor*{Z}{_1_3^0}$ and $\tensor*{Z}{_3_1^0}$, a contribution which can only be included in the more advanced numerical relativity code that allows variations along two spatial directions.  The solid line in Fig.~\ref{fig:1}b is the state space orbit, and  Fig.~\ref{fig:2} shows the initial state of the $\Omega_i$ and representative time steps during the evolution.  

The two figures each show  that the added off-diagonal shear component activated by allowing variations along two (or more) spatial directions qualitatively changes the outcome despite its small amplitude.  The spacetime now rapidly evolves to the flat FRW slow contraction attractor solution, as can be  further verified by checking that the equation of state converges to the attractor value $\varepsilon=\varepsilon^{\ast}=13$. 
Increasing the amplitude or including several $y$-dependent fourier modes to $\tensor*{Z}{_2_3^0}$ does not change the outcome.  

\begin{figure*}[h!]
\begin{center}
\includegraphics[width=7.00in,angle=-0]{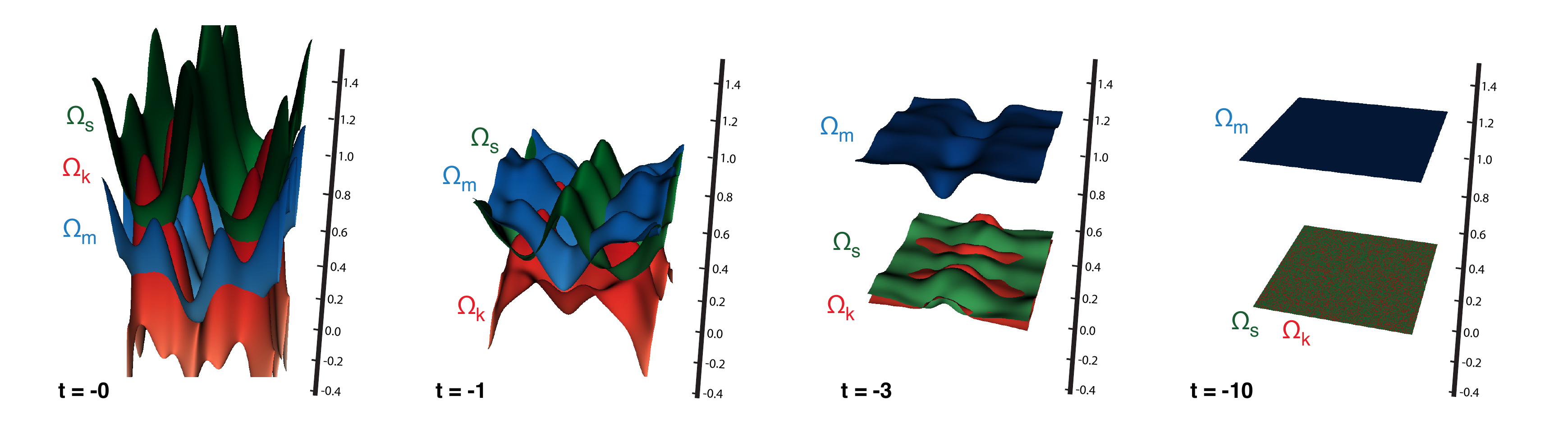}
\end{center}
\caption{The evolution of the relative contributions of the $\Omega_i$ for a representative case with initial conditions that combine multi-mode spatial variations in both the scalar field velocity and the components of  $\tensor*{Z}{_a_b^0} $ in Eq.~ (\ref{ZZ}) as detailed in the text.}   
\label{fig:3}
\end{figure*} 

The example shows that including variations along two (or more) spatial directions and activating more shear components, the rapidity and robustness of smoothing can be dramatically {\it enhanced}.  Conversely, for initial conditions  that smooth the flat FRW when the initial variations are only along one spatial direction, we found no examples where enhancements of the shear  along two dimensions drove the universe to a Kasner-like state.  In the {\it Discussion}, we conjecture why this might occur.  

{\bf Combinations of initial multi-mode variations in the scalar field velocity and shear.}  The final set of tests was for initial conditions that combine spatial variations with multiple fourier modes in both the scalar field initial velocity distribution and all shear components.  Even if the effects of the scalar field velocity  or shear spatial variations alone do not inhibit the rapidity and robustness of smoothing, it is conceivable that non-linear interactions when both are included could.  However, no such effect was found. 
{ 
A representative example with reduced symmetry  is 
 $b_1 =- 0.18$, $b_2 = 0.12$, $a_1 = 0.23$, $a_2 = 0.27$, $\tilde{a}_2 = 0.12$, $c_1 = 0.26$, $c_2 = 0.18$, $\xi=0.02$, $\kappa_1= 0.01$. $\kappa_2=0.02$, 
and $Z_{23}^0 (\vec{x},\, t_0)= Z_{32}^0 (\vec{x},\, t_0)=\kappa_1 + a_2 {\rm cos} \, x + \tilde{a}_2 {\rm cos} \, 3x$ and $Q_0=0.3483$, $q_x=0.1$, $\tilde{q}_x=0.9$,
$q_y=0.11$, $\tilde{q}_y=0.08$ and $Q(\vec{x},\, t_0)= Q_0+ q_x  {\rm cos} \, x + \tilde{q}_x {\rm cos} \, 3x+ q_y {\rm cos} \, 3y +\tilde{q}_y {\rm cos} \, 2x$.  
The dotted line in Fig.~\ref{fig:1}b shows the state space  orbit and Fig.~\ref{fig:3} shows the initial state of the $\Omega_i$ and representative time steps during the evolution.   The rapidity of the smoothing (by $t\approx -10$) is greatly increased compared to the case in Fig.~\ref{fig:2} despite the fact that the initial spatial variations include more fourier modes in $Q(\vec{x},\, t_0)$ and $Z_{ab}^0(\vec{x},\, t_0)$, demonstrating that the variations enhance rapidity.}

\vspace{0.1in}
\noindent
{\it Discussion.}    Earlier numerical relativity simulations with initial conditions entailing a single spatially varying fourier mode and/or variations along a single spatial direction indicated that slow contraction is a remarkably robust and rapid smoother, able to homogenize, isotropize and flatten large deviations from a flat FRW geometry in a matter of a  ten or so $e$-folds of contraction of the Hubble radius (and contraction of the scale factor by less than a factor of two).  A natural question is whether this rapidity and robustness remains when many fourier modes are included and variations along two (or more) spatial directions that reduce the symmetry and activate more shear components are allowed.  This study shows that the answer is a definitive `yes.'  

To conclude, we put forward a conjecture as to why this might occur: In earlier studies, we have shown that, if the stress-energy is supplied by a canonical scalar field minimally coupled to Einstein gravity and with sufficiently steep negative potential energy density,  
\begin{itemize}
\item[-] during contraction the evolution quickly becomes ultralocal, meaning that gradient terms in the evolution equations are negligible compared to velocity terms, see Ref.~\cite{Ijjas:2021gkf};  
\item[-] in the ultralocal limit the basin of attraction of Kasner-like solutions is small  so that for most initial conditions spacetime rapidly evolves towards the flat FRW attractor scaling solution, see Ref.~\cite{Ijjas:2020dws}. 
\end{itemize}

Combining both of these results, we conjecture that the more degrees of freedom are available for the system to deviate from a Kasner-like spacetime, the more initial conditions will evolve towards a flat FRW spacetime and away from the Kasner-like solution.  
Or said another way, given that near-Kasner solutions are unstable, the more degrees of freedom that are active, the more channels that are available to initiate a transition from one Kasner-like epoch to the next, increasing the likelihood
that eventually a transition will land in the basin of attraction of the 
stable, smooth FRW state.
Our conjecture implies that reducing the symmetry assumptions on our initial conditions to include variations in two (or more) spatial directions and/or allowing more modes in each spatial directions enhances the rapidity and robustness of smoothing, meaning that, typically, more initial conditions will evolve towards a flat FRW spacetime. This is especially well illustrated by the examples above.
The results significantly strengthen the case for slow contraction as being a robust and rapid smoother.

 \vspace{0.1in}
\noindent
{\it Acknowledgements.} 
 The work of A.I. is supported by the Lise Meitner Excellence Program of the Max Planck Society and the Simons Foundation grant number 663083.
F.P. acknowledges support from NSF grant PHY-1912171, the Simons Foundation, and the Canadian Institute For Advanced Research (CIFAR).  P.J.S. supported in part by the DOE grant number DEFG02-91ER40671 and by the Simons Foundation grant number 654561.  A.P.S. is also supported by the Simons Foundation grant number 654561.

\bibliographystyle{apsrev}
\bibliography{bib-rapidity}

\begin{thebibliography}{8}
\expandafter\ifx\csname natexlab\endcsname\relax\def\natexlab#1{#1}\fi
\expandafter\ifx\csname bibnamefont\endcsname\relax
  \def\bibnamefont#1{#1}\fi
\expandafter\ifx\csname bibfnamefont\endcsname\relax
  \def\bibfnamefont#1{#1}\fi
\expandafter\ifx\csname citenamefont\endcsname\relax
  \def\citenamefont#1{#1}\fi
\expandafter\ifx\csname url\endcsname\relax
  \def\url#1{\texttt{#1}}\fi
\expandafter\ifx\csname urlprefix\endcsname\relax\def\urlprefix{URL }\fi
\providecommand{\bibinfo}[2]{#2}
\providecommand{\eprint}[2][]{\url{#2}}

\bibitem[{\citenamefont{Khoury et~al.}(2001)\citenamefont{Khoury, Ovrut,
  Steinhardt, and Turok}}]{Khoury:2001wf}
\bibinfo{author}{\bibfnamefont{J.}~\bibnamefont{Khoury}},
  \bibinfo{author}{\bibfnamefont{B.~A.} \bibnamefont{Ovrut}},
  \bibinfo{author}{\bibfnamefont{P.~J.} \bibnamefont{Steinhardt}},
  \bibnamefont{and} \bibinfo{author}{\bibfnamefont{N.}~\bibnamefont{Turok}},
  \bibinfo{journal}{Phys. Rev.} \textbf{\bibinfo{volume}{D64}},
  \bibinfo{pages}{123522} (\bibinfo{year}{2001}), \eprint{hep-th/0103239}.

\bibitem[{\citenamefont{Erickson et~al.}(2007)\citenamefont{Erickson, Gratton,
  Steinhardt, and Turok}}]{Erickson:2006wc}
\bibinfo{author}{\bibfnamefont{J.~K.} \bibnamefont{Erickson}},
  \bibinfo{author}{\bibfnamefont{S.}~\bibnamefont{Gratton}},
  \bibinfo{author}{\bibfnamefont{P.~J.} \bibnamefont{Steinhardt}},
  \bibnamefont{and} \bibinfo{author}{\bibfnamefont{N.}~\bibnamefont{Turok}},
  \bibinfo{journal}{Phys. Rev.} \textbf{\bibinfo{volume}{D75}},
  \bibinfo{pages}{123507} (\bibinfo{year}{2007}), \eprint{hep-th/0607164}.

\bibitem[{\citenamefont{Cook et~al.}(2020)\citenamefont{Cook, Glushchenko,
  Ijjas, Pretorius, and Steinhardt}}]{Cook:2020oaj}
\bibinfo{author}{\bibfnamefont{W.~G.} \bibnamefont{Cook}},
  \bibinfo{author}{\bibfnamefont{I.~A.} \bibnamefont{Glushchenko}},
  \bibinfo{author}{\bibfnamefont{A.}~\bibnamefont{Ijjas}},
  \bibinfo{author}{\bibfnamefont{F.}~\bibnamefont{Pretorius}},
  \bibnamefont{and} \bibinfo{author}{\bibfnamefont{P.~J.}
  \bibnamefont{Steinhardt}}, \bibinfo{journal}{Phys. Lett. B}
  \textbf{\bibinfo{volume}{808}}, \bibinfo{pages}{135690}
  (\bibinfo{year}{2020}), \eprint{2006.01172}.

\bibitem[{\citenamefont{Garfinkle et~al.}(2008)\citenamefont{Garfinkle, Lim,
  Pretorius, and Steinhardt}}]{Garfinkle:2008ei}
\bibinfo{author}{\bibfnamefont{D.}~\bibnamefont{Garfinkle}},
  \bibinfo{author}{\bibfnamefont{W.~C.} \bibnamefont{Lim}},
  \bibinfo{author}{\bibfnamefont{F.}~\bibnamefont{Pretorius}},
  \bibnamefont{and} \bibinfo{author}{\bibfnamefont{P.~J.}
  \bibnamefont{Steinhardt}}, \bibinfo{journal}{Phys. Rev.}
  \textbf{\bibinfo{volume}{D78}}, \bibinfo{pages}{083537}
  (\bibinfo{year}{2008}), \eprint{0808.0542}.

\bibitem[{\citenamefont{Ijjas et~al.}(2020)\citenamefont{Ijjas, Cook,
  Pretorius, Steinhardt, and Davies}}]{Ijjas:2020dws}
\bibinfo{author}{\bibfnamefont{A.}~\bibnamefont{Ijjas}},
  \bibinfo{author}{\bibfnamefont{W.~G.} \bibnamefont{Cook}},
  \bibinfo{author}{\bibfnamefont{F.}~\bibnamefont{Pretorius}},
  \bibinfo{author}{\bibfnamefont{P.~J.} \bibnamefont{Steinhardt}},
  \bibnamefont{and} \bibinfo{author}{\bibfnamefont{E.~Y.}
  \bibnamefont{Davies}}, \bibinfo{journal}{JCAP} \textbf{\bibinfo{volume}{08}},
  \bibinfo{pages}{030} (\bibinfo{year}{2020}), \eprint{2006.04999}.

\bibitem[{\citenamefont{Ijjas et~al.}(2021)\citenamefont{Ijjas, Sullivan,
  Pretorius, Steinhardt, and Cook}}]{Ijjas:2021gkf}
\bibinfo{author}{\bibfnamefont{A.}~\bibnamefont{Ijjas}},
  \bibinfo{author}{\bibfnamefont{A.~P.} \bibnamefont{Sullivan}},
  \bibinfo{author}{\bibfnamefont{F.}~\bibnamefont{Pretorius}},
  \bibinfo{author}{\bibfnamefont{P.~J.} \bibnamefont{Steinhardt}},
  \bibnamefont{and} \bibinfo{author}{\bibfnamefont{W.~G.} \bibnamefont{Cook}}
  (\bibinfo{year}{2021}), \eprint{2103.00584}.

\bibitem[{\citenamefont{York}(1971)}]{York:1971hw}
\bibinfo{author}{\bibfnamefont{J.}~\bibnamefont{York},
  \bibfnamefont{James~W.}}, \bibinfo{journal}{Phys. Rev. Lett.}
  \textbf{\bibinfo{volume}{26}}, \bibinfo{pages}{1656} (\bibinfo{year}{1971}).

\bibitem[{\citenamefont{Ijjas and Steinhardt}(2019)}]{Ijjas:2019pyf}
\bibinfo{author}{\bibfnamefont{A.}~\bibnamefont{Ijjas}} \bibnamefont{and}
  \bibinfo{author}{\bibfnamefont{P.~J.} \bibnamefont{Steinhardt}},
  \bibinfo{journal}{Phys.Lett.} \textbf{\bibinfo{volume}{B795}},
  \bibinfo{pages}{666} (\bibinfo{year}{2019}), \eprint{1904.08022}.

\end{thebibliography}

\end{document}